\documentclass[journal]{IEEEtran} %Journal
\usepackage{cite}

\ifCLASSINFOpdf
  \usepackage[pdftex]{graphicx}

  \DeclareGraphicsExtensions{.pdf,.jpeg,.png}
\else
   \usepackage[dvips]{graphicx}

   \DeclareGraphicsExtensions{.eps}
\fi

\usepackage[cmex10]{amsmath}

\usepackage{array}
\usepackage{mdwmath}
\usepackage{mdwtab}

\usepackage{eqparbox}

\usepackage[caption=false,font=footnotesize]{subfig}

\usepackage{multirow}
\usepackage{amsfonts}
\usepackage{rotating}

\newcommand{\var}{{\mathbb V}\text{ar}}
\newcommand{\sign}{\text{sign}}
\newcommand{\esp}{\mathbb E}
\newcommand{\expon}{\text{e}}

\begin{document}
%
% paper title
% can use linebreaks \\ within to get better formatting as desired
\title{Adaptive Quantizers for Estimation}
%
%
% author names and IEEE memberships
% note positions of commas and nonbreaking spaces ( ~ ) LaTeX will not break
% a structure at a ~ so this keeps an author's name from being broken across
% two lines.
% use \thanks{} to gain access to the first footnote area
% a separate \thanks must be used for each paragraph as LaTeX2e's \thanks
% was not built to handle multiple paragraphs
%
\author{Rodrigo~Cabral Farias~% 
             and~Jean-Marc~Brossier% <-this % stops a space
\thanks{The authors are with Grenoble Laboratory of Images, Speech, Signal and Automatics, Department of Images and Signals, 38402 Saint Martin d'H\`eres, France (e-mail: rodrigo.cabral-farias@gipsa-lab.grenoble-inp.fr;	  jean-marc.brossier@gipsa-lab.grenoble-inp.fr).}}% <-this % stops a space
%\thanks{Manuscript received January XX, XXXX; revised January XX, XXXX.}}
% The paper headers
%\markboth{IEEE TRANS. ON SIGNAL PROCESSING,~Vol.~XX, No.~XX, January~20XX}%
%{Shell \MakeLowercase{FARIAS AND BROSSIER: ADAPTIVE QUANTIZERS FOR ESTIMATION}}
% The only time the second header will appear is for the odd numbered pages
% after the title page when using the twoside option.
% 
% *** Note that you probably will NOT want to include the author's ***
% *** name in the headers of peer review papers.                   ***
% You can use \ifCLASSOPTIONpeerreview for conditional compilation here if
% you desire.

% If you want to put a publisher's ID mark on the page you can do it like
% this:
%\IEEEpubid{0000--0000/00\$00.00~\copyright~20XX IEEE}
% Remember, if you use this you must call \IEEEpubidadjcol in the second
% column for its text to clear the IEEEpubid mark.

% use for special paper notices
%\IEEEspecialpapernotice{(Invited Paper)}

% make the title area
\maketitle

\begin{abstract}
In this paper, adaptive estimation based on noisy quantized observations is studied. A low complexity adaptive algorithm using a quantizer with adjustable input gain and offset is presented. Three possible scalar models for the parameter to be estimated are considered: constant, Wiener process and Wiener process with deterministic drift. After showing that the algorithm is asymptotically unbiased for estimating a constant, it is shown, in the three cases, that the asymptotic mean squared error depends on the Fisher information for the quantized measurements. It is also shown that the loss of performance due to quantization depends approximately on the ratio of the Fisher information for quantized and continuous measurements. At the end of the paper the theoretical results are validated through simulation under two different classes of noise, generalized Gaussian noise and Student's-t noise.
\end{abstract}

% Note that keywords are not normally used for peerreview papers.

% For peer review papers, you can put extra information on the cover
% page as needed:
 %\ifCLASSOPTIONpeerreview
 %\begin{center} \bfseries EDICS Category: SSP-PARE \end{center}
 %\fi
%
% For peerreview papers, this IEEEtran command inserts a page break and
% creates the second title. It will be ignored for other modes.
\IEEEpeerreviewmaketitle

%%%%%%%%%%%%%%%%%%%%%%%%%%%%%%%%%%%%%%%%%%%%%%%%%%%%%%%%%%%

\section{Introduction}
\IEEEPARstart{C}{ontinuous} advances in the development of cheaper and smaller sensors and communication devices motivated the introduction of sensor networks in many different domains, \textit{e.g.} military applications, infrastructure security, environment monitoring, industrial applications and traffic monitoring \cite{Chong2003}. When designing a sensing system, one must account not only for the physical perturbations that can affect sensing performance,  more specifically noise, but also for the inherent design constraints such as bandwidth and complexity limitations. Commonly, the effect of the noise in system performance is taken into account, but bandwidth and complexity constraints are neglected.

One simple way to respect bandwidth constraints is to compress sensor information using quantizers. The theory of quantizer design for reducing distortion in the measurement representation is well established in the literature \cite{Gersho1992}, however much less results can be found when the quantities to be reconstructed are not directly the measurements but an underlying parameter embedded in noise.

In \cite{Papadopoulos2001}, noisy samples of a constant are taken using a uniform quantizer with an input offset, the output samples of the quantizer are used to estimate the constant. Using this type of measurement system, results for different types of offset were obtained. The types of offset considered were known constant and variable offset, random offset and offset based on feedback of the output measurements. The comparison was performed based on the Cram\'er--Rao bound (CRB) ratio which is the worst case ratio between the CRB for quantized measurements and continuous measurements. It was shown that the last type of offset, based on feedback, was the most efficient one.

Another interesting result from \cite{Papadopoulos2001} is that in the Gaussian noise case with one bit quantized measurements, the minimum CRB ratio that can be attained is $ \frac{\pi}{2} $. This result was used as a motivation for \cite{Ribeiro2006a} to study more in detail estimation under Gaussian noise and binary quantization. In \cite{Ribeiro2006a}, it was shown that the CRB for a fixed known threshold can be upper bounded by the exponential of the squared difference between the threshold and the constant to be estimated. This means that the closer the threshold is to the parameter to be estimated with binary measurements, the lower can be the estimation variance. It was also pointed out that an iterative algorithm could be used to adjust the threshold exactly to be the last estimate of the parameter.

An adaptive algorithm for placing the threshold was detailed in \cite{Li2007}, where a sensor network extension was also proposed. At each time step, a sensor measures one bit, updates its threshold using a simple cumulative sum and broadcasts the new threshold to the other sensors and to a fusion center. Thus, the thresholds are placed around the parameter in an adaptive way and at the fusion center the broadcasted bits are used to obtain a more precise estimate of the parameter. Two other methods for updating the thresholds were presented in \cite{Fang2008}, one method used a more refined cumulative sum  based on the last two measured bits, the other proposed method was to estimate the parameter using a maximum likelihood method and then set the threshold at the estimate of the parameter. It was shown that in the asymptotic case (large number of iterates) the CRB for the fusion center estimate using maximum likelihood threshold updates converges to the minimum possible CRB, which is the CRB when the threshold is placed exactly at the parameter.

In the same line of the work mentioned above, algorithms for estimating a scalar parameter from multiple bit quantized noisy measurements are proposed. The algorithms developed in this work are based on low complexity adaptive techniques that can be easily implemented in practice. The mean and mean squared error (MSE) are obtained for a general class of symmetrically distributed noise and three types of parameter evolution: constant, Wiener process and Wiener process with drift. As in related work \cite{Papadopoulos2001}, the loss of estimation performance due to quantization is also evaluated and the validity of the performance results is verified through simulation.

The main contributions of this work are

\begin{itemize}
\item \emph{Design and analysis of adaptive estimation algorithms based on multiple bit quantized noisy measurements.} Differently of \cite{Li2007} and \cite{Fang2008}, where only binary quantization is treated.
\newpage
\item \emph{Explicit performance analysis for tracking of a varying parameter.} In \cite{Papadopoulos2001,Ribeiro2006a,Li2007,Fang2008} the parameter is set to be constant and all subsequent analysis is based on this hypothesis.
\item \emph{Low complexity algorithms.} The algorithms proposed here are based on simple recursive techniques that have lower complexity than the maximum likelihood methods used in \cite{Li2007} and \cite{Fang2008}.
\end{itemize}

The paper is structured in the following form: in section II the problem is stated and the main assumptions are made, in section III the general adaptive algorithm and results from adaptive algorithms theory are presented, then in section IV the parameters of the adaptive algorithm are obtained. Section V contains theoretical performance results and also the simulation of the algorithm. Section VI concludes the paper.

%%%%%%%%%%%%% First part %%%%%%%%%%%%%
\section{Problem statement}
Let $ \mathbf{X} $ be a stochastic process defined on the probability space $ \mathcal{P}=\left( \Omega,\mathcal{F},\mathbb{P}\right)  $ with values on $ \left( \mathbb{R},\mathcal{B}\left( \mathbb{R}\right)   \right)  $, at each instant $ k \in \mathbb{N}^{\star} $, the corresponding scalar random variable (r.v.) $ X_{k} $ will be given by the following model:

\begin{equation}\label{eq1}
X_{k}=X_{k-1}+W_{k}, 
\end{equation}

\noindent where $ W_{k} $ is a sequence of independent Gaussian random variables with its mean given by a small amplitude deterministic unknown sequence $ u_{k} $ and small known standard deviation $ \sigma_{w} $:
\begin{equation}\label{eqw}
W_{k}\sim \mathcal{N}\left(u_{k},\sigma_{w}^{2} \right).
\end{equation}

\noindent The initial condition $ X_{0} $ will be considered to be an unknown deterministic constant.

The model expressed in (\ref{eq1}) is a compact form to describe three different evolution models for $ X_{k} $:

\begin{itemize}
\item \emph{Constant}: by taking $ u_{k}=\sigma_{w}=0 $, then $ X_{k}=X_{0}=x $ is an unknown deterministic constant.
\item \emph{Wiener process}: if $ u_{k}=0 $, $ \sigma_{w}>0 $ and small , then $ X_{k} $ is a slowly varying Wiener process. This model is commonly used to describe a slowly varying parameter of a system when the model for its evolution is random but with unknown form.
\item \emph{Wiener process with drift}: in this case $ u_{k} $ and $ \sigma_{w} $ are non zero and with small amplitudes. The fact that $ u_{k} $ is nonzero makes the Wiener process to have a drift, thus representing a model with a deterministic component that is perturbed by small random fluctuations.
\end{itemize}

 The process $ \mathbf{X} $ is observed through $ \mathbf{Y} $ and they are related as follows:
 
 \begin{equation}\label{eq2}
Y_{k}=X_{k}+V_{k}, 
\end{equation}

\noindent where the noise $ V_{k} $ is a sequence of additive independent and identically distributed (i.i.d.) r.v. which is also independent of $ W_{k} $. The cumulative distribution function (CDF) of $ V_{k} $ will be denoted by $ F $. Some assumptions on $ F $ are stated below.

\emph{Assumptions (on the noise distribution)}: 
\begin{itemize}
\item [A1. ]$ F $ is locally Lipschitz continuous.
\item [A2. ]$ F $ admits a probability density function (PDF) $ f $ with respect to (w.r.t.) the standard Lebesgue measure on $ \left( \mathbb{R},\mathcal{B}\left( \mathbb{R}\right)   \right)  $.
\item [A3. ]The PDF  $ f\left(x\right)  $ is an even function and it strictly decreases w.r.t. $ \vert x \vert $.
\end{itemize}

The first assumption is required by the method of analysis that will be used to assess the performance of the proposed algorithms. Most noise CDFs considered in practice are Lipschitz continuous, thus the first assumption is generally satisfied. Assumption 2 is a commonly used assumption that in practice will be used when the derivative of $ F $ w.r.t. its arguments is needed. Assumption 3 will be used to prove the asymptotic convergence of the algorithms and it is also commonly satisfied in practice.

The observations are quantized using an adjustable quantizer whose output is given by 

\begin{equation}\label{eq3}
i_{k}=Q\left( \frac{Y_{k}-b_{k}}{\Delta_{k}}\right) , 
\end{equation}

\noindent where $ i_{k} $ is an integer defined on a finite set of $ N_{I} $ integers, $ N_{I} $ being the number of quantization intervals. The  quantizer parameters $ b_{k} $ and $ \frac{1}{\Delta_{k}} $ are sequences of adjustable offsets and gains respectively. The function $ Q $ represents a static normalized quantizer and it is characterized by $ N_{I}+1 $ thresholds. For simplification purposes some assumptions on the quantizer will be used.

\emph{Assumptions (on the quantizer)}: 
\begin{itemize}
\item [A4. ]$ N_{I} $ will be considered to be an even natural number and
\begin{equation}
 i_{k}\in I=\left\lbrace -\frac{N_{I}}{2} ,\ldots,-1,+1,\ldots,+\frac{N_{I}}{2}\right\rbrace.  \nonumber
\end{equation}
\item [A5. ] It will be assumed that the static quantizer is symmetric and centered at zero. This means that the vector of thresholds\footnote{Infinite thresholds are used to have the same notation for the probabilities of the granular and overload regions.}
\begin{equation}
\boldsymbol{\tau}= \left[  \tau_{\footnotesize{-\frac{N_{I}}{2}}} \;\ldots\,\tau_{-1} \;\tau_{0}\; \tau_{1} \;\ldots\; \tau_{ \frac{N_{I}}{2}} \right]^{T} \nonumber
\end{equation}
has elements given by the following expressions
\begin{eqnarray}\label{eq4}
&\tau_{0}&=0, \nonumber \\ &\tau_{i}&=-\tau_{-i}, \quad \forall i \in  \left\lbrace 1,\cdots ,\frac{N_{I}}{2}\right\rbrace,\nonumber \\ &\tau_{\frac{N_{I}}{2}}&=+\infty .
\end{eqnarray}
\end{itemize}

These assumptions will be used later to simplify the choice of parameters of the algorithms.

For $ \frac{\left\vert Y_{k}-b_{k} \right\vert}{\Delta_{k}}\in\left[ \tau_{i-1},\tau_{i}\right) $, the adjustable quantizer output is given by
\begin{eqnarray}\label{eq5}
i_{k}=Q\left(\frac{Y_{k}-b_{k}}{\Delta_{k}}\right)=i\,\mathrm{sign}\left( Y_{k}-b_{k}\right).  
\end{eqnarray}

\begin{figure}[!t]
\centering
\includegraphics[width=0.7\linewidth ]{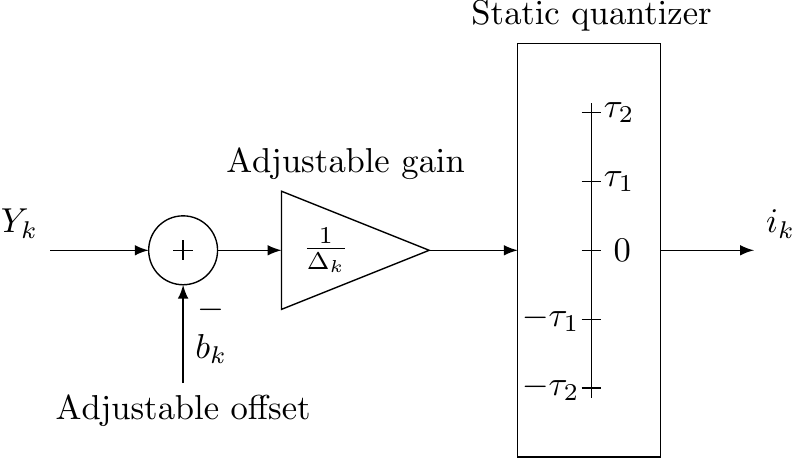}
\caption{Scheme representing the adjustable quantizer. The offset and gain can be adjusted dynamically while the quantizer thresholds are fixed.}
\label{fig1}
\end{figure}

A scheme representing the quantizer is given in Fig. \ref{fig1}. Note that even if the quantizer is not uniform (with constant distance between thresholds), it can be implemented using a uniform quantizer with a compander approach \cite{Gersho1992}.

Based on the quantizer outputs the main objective is to estimate $ X_{k} $ and a  secondary objective is to adjust the parameters $ b_{k} $ and $ \Delta_{k} $ to enhance estimation performance. As the estimate $ \hat{X}_{k} $ of $ X_{k} $ will be possibly used in real time applications, it might be estimated online, which means that $ \hat{X}_{k} $ will only depend on past and present $ i_{k} $.  To simplify it will be considered that the offset is set to be $ \hat{X}_{k-1} $ and that the gain is set to be a constant $ \Delta $. For the adaptive algorithm presented later, the fact that the offset is set to $ \hat{X}_{k-1} $ will have, as a consequence, an asymptotic performance that does not depend on the mean of $ X_{k} $, thus simplifying the analysis. The choice of $ \Delta $ is discussed in section IV.

\begin{figure}[!t]
\centering
\includegraphics[width=1\linewidth ]{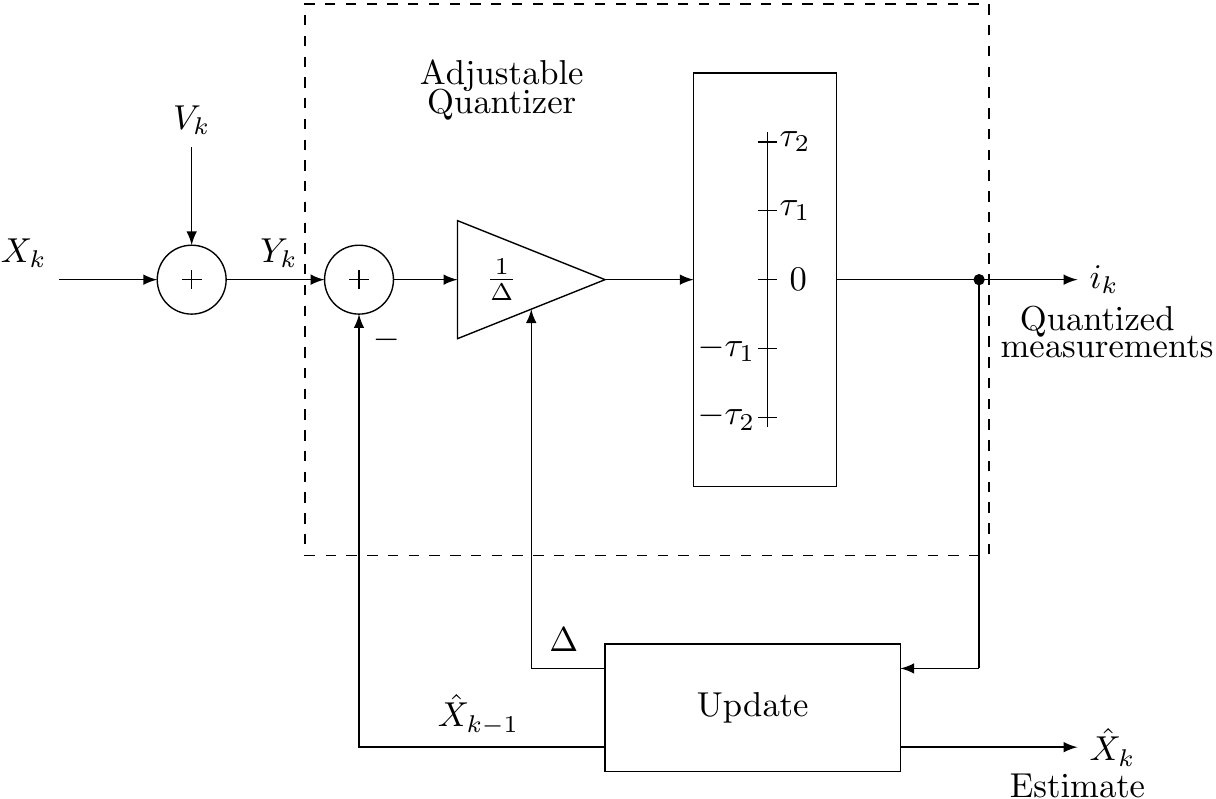}
\caption{Block representation of the estimation scheme. The estimation algorithm and the procedures to set the offset and the gain are represented by the \textit{Update} block.}
\label{fig2}
\end{figure}

The general scheme for the estimation of $ X_{k} $ is depicted in Fig. \ref{fig2} and the main objective will be to find a low complexity algorithm that will be placed in the block named \textit{Update}.

\section{General algorithm}

A simple and general form for the estimation algorithm that respects the constraints defined above (low complexity and online) is the following adaptive algorithm:

\begin{equation}\label{eq6}
\hat{X}_{k}=\hat{X}_{k-1}+\gamma_{k}\eta\left[Q\left(\frac{Y_{k}-\hat{X}_{k-1}}{\Delta}\right) \right]  . 
\end{equation}

\noindent In the expression above, $ \gamma_{k} $ is a sequence of positive real gains and $ \eta[\cdot] $ is a mapping from $ I $ to $ \mathbb{R} $ that is defined as a sequence of $ N_{I} $ coefficients $\left\lbrace \eta_{-\frac{N_{I}}{2}},\ldots,\eta_{-1},\eta_{1},\ldots,\eta_{\frac{N_{I}}{2}} \right\rbrace$, these coefficients are equivalent to the output quantization levels used in quantization theory. The use of this algorithm is also motivated by the following observations:

\begin{itemize}
\item when estimating a constant, the maximum likelihood estimator can be approximated by a simpler online algorithm using a stochastic gradient ascent algorithm, which has the same form as (\ref{eq6}). It is shown in section IV that for the optimal choice of $ \eta_{i} $, (\ref{eq6}) is equivalent to a stochastic gradient ascent method to maximize the log-likelihood.
\item To estimate a Wiener process, a simple choice of estimator is a Kalman filter like method based on the quantized innovation, which is also (\ref{eq6}).
\end{itemize}

Due to the symmetry of the noise distribution, when $ \hat{X}_{k} $ is close to $ X_{k} $, it seems reasonable to suppose that the corrections given by the output quantizer levels have odd symmetry with positive values for positive $ i $, this symmetry will be useful later for simplification purposes. Thus, one assumption will be added to A1-A5.

\emph{Assumption (on the quantizer output levels)}: 
\begin{itemize}
\item [A6. ]The quantizer output levels have odd symmetry w.r.t. $ i $:
\begin{equation}\label{oddsymm}
\eta_{i}=-\eta_{-i},
\end{equation}
\noindent with $ \eta_{i}>0 $ for $ i>0 $.
\end{itemize}

The non differentiable non linearity in (\ref{eq6}) makes it difficult to be analyzed. Fortunately, an analysis based on mean approximations was developed in \cite{Benveniste1990} for a wide class of adaptive algorithms, within this framework, the function $ \eta $ could be a general non linear non differentiable function of $ Y_{k} $ and $ \hat{X}_{k} $ and it was shown that the gains $ \gamma_{k} $ that optimizes the estimation of $ X_{k} $ should be as follows:

\begin{itemize}
\item $ \gamma_{k}\propto\frac{1}{k}$ when $ X_{k} $ is constant.
\item $ \gamma_{k} $ is constant for a Wiener process $ X_{k} $.
\item $ \gamma_{k}\propto u_{k}^{\frac{2}{3}} $ when $ X_{k} $ is a Wiener process with drift.
\end{itemize}

In the following parts of this section the results of \cite{Benveniste1990} will be applied for the analysis of (\ref{eq6}) in the three evolution models of $ X_{k} $.

\subsection{Constant $ X_{k} $}
In this case $ X_{k}=x $. To obtain convergence of $ \hat{x}_{k} $ to a constant, the gains must be:

 \begin{equation}\label{eq7}
\gamma_{k}=\frac{\gamma}{k}.
\end{equation}

For large $ k $, the mean trajectory of $ \hat{X}_{k} $ can be approximated using the ordinary differential equation (ODE) method. The ODE method approximates the expectation of the estimator $ \esp\left[  \hat{X}_{k} \right]  $ by $ \hat{x}\left(t_{k}\right)  $, where $ \hat{x}\left(t\right)  $ is the solution of

\begin{equation}\label{eq8}
\frac{d\hat{x}}{dt}=\gamma h\left(\hat{x}\right) , 
\end{equation}
\noindent the correspondence between continuous and discrete time is given by $ t_{k}=\sum\limits_{j=1}^{k} \frac{1}{j} $ and $ h\left(\hat{x}\right) $ is the following:
\begin{equation}\label{eq9}
h\left(\hat{x}\right)=\esp\left[\eta\left(Q\left( \frac{x-\hat{x}+V }{\Delta }\right)  \right)  \right]  , 
\end{equation}

\noindent where the expectation is evaluated w.r.t. $ F\left(v\right)  $.

For the solution of (\ref{eq8}) to be valid as an approximation of $ \esp\left[  \hat{X}_{k} \right]  $, $ h\left(\hat{x}\right)  $ has to be a locally Lipschitz continuous function of $ \hat{x} $. Using the assumptions on the quantizer thresholds and output levels, the expectation in (\ref{eq9}) can be written as:

\begin{equation}\label{eq10}
h\left(\hat{x}\right)=\sum\limits_{i=1}^{\frac{N_{I}}{2}} \left[  \eta_{i} F_{d}\left(i,\hat{x},x\right)-\eta_{i} F_{d}\left(-i,\hat{x},x\right)\right]   , 
\end{equation}

\noindent where $ F_{d} $ is a difference of CDFs:

\begin{equation}\label{eq11}
 F_{d}=
 \begin{cases}
F\left(\tau_{i}\Delta+ \hat{x} -x \right) -F\left(\tau_{i-1}\Delta+ \hat{x} -x \right) \\ \qquad\qquad\qquad\qquad\qquad\qquad
\text{if $ i \in \left\lbrace 1,\cdots,\frac{N_{I}}{2} \right\rbrace$},\\ 
F\left(\tau_{i+1}\Delta+\hat{x} -x \right) -F\left(\tau_{i}\Delta+ \hat{x}-x\right) \\ \qquad\qquad\qquad\qquad\qquad\qquad \text{if $i \in \left\lbrace -1,\cdots,-\frac{N_{I}}{2} \right\rbrace $}.\\ 
 \end{cases} 
\end{equation}

From assumption A1, the function $ h $ is a linear combination of locally Lipschitz continuous functions, which implies that $ h $ is also locally Lipschitz continuous, thus the ODE method can be applied.

If $ \hat{x}\rightarrow x $ when $ t\rightarrow \infty $ for all $ x $ and all $ \hat{x}\left(0\right)  $, the adaptive algorithm is asymptotically unbiased, and in this case it can also be shown, using a central limit theorem, that the estimation error is asymptotically distributed as a Gaussian r.v. \cite[pp. 109]{Benveniste1990}:

\begin{equation}\label{eq12}
\sqrt{k} \left(\hat{X}_{k}-x\right)  \underset{k\rightarrow\infty} \rightsquigarrow \mathcal{N}\left(0,\sigma_{\infty}^{2} \right),
\end{equation}

\noindent where the asymptotic variance $ \sigma_{\infty}^{2} $ is given by:

\begin{equation}\label{eq13}
\sigma_{\infty}^{2}=\frac{\gamma^{2}R\left(x\right) }{-2\gamma h_{\hat{x}}\left(x\right)-1 },
\end{equation}

\begin{itemize}
\item The term denoted $ R $ in the numerator is the variance of the adaptive algorithm normalized increments $ \left( \frac{\hat{X}_{k}-\hat{X}_{k-1}}{\gamma_{k}}\right)  $ when $ \hat{x} $ is equal to $ x $. From A3 and A6, $ h\left(\hat{x}\right)=0 $ when $ \hat{x}=x $ and this variance can be written as the second order moment of the quantizer output levels:

\begin{eqnarray}\label{eq14}
R\left(x\right)&=&\left.\var\left[ \eta\left(Q\left(\frac{x-\hat{x}+V}{\Delta}\right) \right)\right]\right\vert_{\hat{x}=x}\nonumber\\ 
&=&\sum\limits_{i=1}^{\frac{N_{I}}{2}} \left( \eta_{i}^2 F_{d}\left(i,x,x\right)+\eta_{-i}^2 F_{d}\left(-i,x,x\right) \right)\nonumber \\
&=&2\sum\limits_{i=1}^{\frac{N_{I}}{2}}  \eta_{i}^2 F_{d}\left(i,x,x\right),
\end{eqnarray}
\noindent where the last equality comes from the symmetry assumptions.

\item The term in the denominator is the derivative of $ h $ when $ \hat{x} $ is equal to $ x $:

\begin{eqnarray}\label{eq15}
 h_{\hat{x}}\left(x\right) &=& \left. \frac{dh}{d\hat{x}}\right\vert_{\hat{x}=x}\\ \nonumber
 &=& -\sum\limits_{i=1}^{\frac{N_{I}}{2}}  \left[  \eta_{i} f_{d}\left(i,x,x\right)-\eta_{i} f_{d}\left(-i,x,x\right)\right] ,
\end{eqnarray}

with

\begin{equation}\label{eq16}
 f_{d}=
 \begin{cases}
f\left(\tau_{i-1}\Delta+ \hat{x} -x \right) -f\left(\tau_{i}\Delta+ \hat{x} -x \right) \\ \quad\qquad\qquad\qquad\qquad \text{if $i \in \left\lbrace 1,\cdots,\frac{N_{I}}{2} \right\rbrace$},\\ 
f\left(\tau_{i}\Delta+ \hat{x} -x \right) -f\left(\tau_{i+1}\Delta+ \hat{x} -x \right) \\ \quad\qquad\qquad\qquad\qquad\text{if $ i \in \left\lbrace -1,\cdots,-\frac{N_{I}}{2} \right\rbrace$}.\\ 
 \end{cases} 
\end{equation}

\noindent From the symmetry assumptions, $ f_{d}\left(i,x,x\right)  $ is odd w.r.t. $ i $, thus (\ref{eq15}) can be rewritten as

\begin{equation}\label{eqhx}
h_{\hat{x}}\left(x\right)=-2\sum\limits_{i=1}^{\frac{N_{I}}{2}}  \eta_{i} f_{d}\left(i,x,x\right).
\end{equation}
\end{itemize}
Minimizing $ \sigma_{\infty}^{2} $ w.r.t. the positive gain $ \gamma $ gives 

\begin{equation}\label{eq17}
\gamma^{\star}=-\frac{1}{h_{\hat{x}}\left(x\right)}
\end{equation}
\begin{equation}\label{eq18}
\sigma_{\infty}^{2}=\frac{R\left(x\right)}{h_{\hat{x}}^{2}\left(x\right)}.
\end{equation}

When $ \hat{x}=x $, the functions $ F_{d}\left(i,\hat{x},x\right) $ and $ f_{d}\left(i,\hat{x},x\right) $ do not depend on $ x $ anymore, thus from now on they will be denoted $ F_{d}\left[i\right]  $ and $ f_{d}\left[i\right]  $. The functions $ R\left(x\right)  $ and $ h_{\hat{x}}\left(x\right) $ do not depend on $ x $ either, thus they will be denoted by the constants $ R $ and $ h_{\hat{x}} $ respectively.

To specify completely the adaptive algorithm, the quantizer parameters $ \eta_{i} $, $ \boldsymbol{\tau} $ and $ \Delta $ can be chosen to minimize (\ref{eq18}).

\subsection{Wiener process}

If $ X_{k} $ is a Wiener process, the mean of $ W_{k} $ is $ u_{k}=0 $ and the variance is a known constant $\var\left[W_{k}\right]=\sigma_{w}^{2} $.  The algorithm gain can be chosen to be a constant $ \gamma_{k}=\gamma $. For small $ \sigma_{w}^{2} $, the mean trajectory of $ \hat{X}_{k} $ is also approximated by (\ref{eq8}), $ x $ being the initial condition $ x_{0} $ of the Wiener process, which is equal to its mean for every $ k $. Thus, if $ \hat{x} $ converges to $ x $, the algorithm is asymptotically unbiased and, in this case, it can be shown that the asymptotic estimation MSE can be approximated in the following way \cite[pp. 130-131]{Benveniste1990}:

\begin{equation}\label{eq19}
\mathrm{MSE}_{\infty}=\underset{k\rightarrow \infty}{\lim} \esp \left[\hat{X}_{k}-X_{k} \right]^{2} \approx \gamma \esp \left[\xi_{t} \right]^{2}.
\end{equation}

The stochastic process $ \xi_{t} $ is the solution of a stochastic differential equation:

\begin{equation}\label{eq20}
d\xi_{t} = h_{\hat{x}} \xi_{t} dt - \gamma \sigma_{w}\sqrt{R} dZ_{t} ,
\end{equation}

\noindent where $ Z_{t} $ is a continuous time Wiener process with unit increment variance. Under the condition

\begin{equation}\label{eq20}
\gamma h_{\hat{x}}<0 ,
\end{equation}

\noindent $\xi_{t} $ is stationary with a marginal Gaussian density $ \mathcal{N}\left(0,\sigma_{\xi}^{2}\right)  $, where the variance is

\begin{equation}\label{eq21}
\sigma_{\xi}^{2}=\frac{\gamma^{2} R+\sigma_{w}^{2}}{-2\gamma h_{\hat{x}}}.
\end{equation}

\noindent Thus, $ \text{MSE}_{\infty} $ can be approximated by $ \sigma_{\xi}^{2} $. Minimizing $ \text{MSE}_{\infty} $ w.r.t. $ \gamma $ gives the optimal $ \gamma $

\begin{equation}\label{eq22}
\gamma^{\star}=\frac{\sigma_{w}}{\sqrt{ R }},
\end{equation}

\noindent which is a positive real, thus changing the condition (\ref{eq20}) into

\begin{equation}\label{eq23}
h_{\hat{x}}<0.
\end{equation}

The MSE for $ \gamma^{\star} $ is

\begin{equation}\label{eq24}
\mathrm{MSE}_{\infty}=\frac{\sigma_{w}\sqrt{R}}{-h_{\hat{x}}}.
\end{equation}

\noindent Using (\ref{eq24}) and (\ref{eq18}) the MSE can be rewritten as

\begin{equation}\label{eq25}
\mathrm{MSE}_{\infty}=\sigma_{w}\sigma_{\infty}.
\end{equation}

\noindent Both the asymptotic MSE for estimating a Wiener process and the asymptotic variance for estimating a constant depend on the quantizer parameters through $ \sigma_{\infty} $, therefore the optimal quantizer parameters will be the same in both cases. The only difference in the adaptive algorithms for these two cases is the sequence of gains $ \gamma_{k} $.

\subsection{Wiener process with drift}

In this case the mean of $ W_{k} $ is nonzero and given by a small amplitude sequence $ u_{k} $, the variance is a constant $ \sigma_{w} $. The gain $ \gamma_{k} $ will be considered to be variable in time and under the assumption of asymptotic unbiasedness for constant $ X_{k} $, the MSE can be approximated by the term due to the estimation bias which is given by \cite[pp. 136]{Benveniste1990}:

\begin{equation}\label{eq26}
\mathrm{MSE}_k=\esp\left[\hat{X}_{k}-X_{k} \right]^{2}\approx \frac{u_{k}^{2}}{\gamma_{k}^{2}h_{\hat{x}}^{2}}-\gamma_{k}\frac{R}{2h_{\hat{x}}}.
\end{equation}

Minimization w.r.t. $ \gamma_{k} $ leads to

\begin{equation}\label{eq27}
\gamma_{k}^{\star}=\left[\frac{4u_{k}^{2}}{-h_{\hat{x}}R}\right] ^{\frac{1}{3}}
\end{equation}
\begin{equation}\label{eq28}
\mathrm{MSE}_{k}\approx 3\left[\frac{u_{k}}{4} \frac{R}{h_{\hat{x}}^{2}} \right]^{\frac{2}{3}}.
\end{equation}

Note that in practice, $ u_{k} $ may be unknown and it will be necessary to replace its value in $ \gamma_{k}^{\star} $ by an estimate of it $ \hat{U}_{k} $, which can be also obtained adaptively, for example by calculating a recursive mean on $ \hat{X}_{k}-\hat{X}_{k-1} $.

The MSE can also be rewritten as a function of $ \sigma_{\infty}^{2} $ with a dependence on $ u_{k} $

\begin{equation}\label{eq29}
\mathrm{MSE}_{k}\approx 3\left[\frac{u_{k}}{4} \sigma_{\infty}^{2} \right]^{\frac{2}{3}}.
\end{equation}

Also in this case the MSE is an increasing function of $ \sigma_{\infty} $. From the three cases it is possible to see that the quantizer design will depend on the following:

\begin{enumerate}
\item \emph{Asymptotic unbiasedness}: it is necessary to prove asymptotic unbiasedness of the algorithm when $ X_{k} $ is constant for the MSE results given above to be valid. This can be done by proving the asymptotic global stability of the ODE (\ref{eq8}) for an arbitrary $ X_{k}=x $ and $ \hat{X}_{0}=\hat{x}\left(0\right)  $ in $ \mathbb{R} $. 

\item \emph{Minimization of $ \sigma_{\infty}^{2} $}: the quantizer parameters can be chosen to minimize $ \sigma_{\infty}^{2} $ and, as a consequence, they will maximize the performance for the three evolution models of $ X_{k} $.
\end{enumerate}

\section{Asymptotic unbiasedness and adaptive algorithm design}

In this section, first it will be shown that the algorithm is asymptotically unbiased. Then, optimization of the algorithm asymptotic performance will be done by minimizing $ \sigma_{\infty}^{2}$ ,which depends on $ \eta_{i} $, $ \Delta\left(x\right)  $ and $ \boldsymbol{\tau} $. The optimal coefficients $ \eta_{i} $ will be found and then the choice for the parameters $ \Delta $ and $ \boldsymbol{\tau} $ will be discussed.

\subsection{Asymptotic unbiasedness}

For the asymptotic performance results to be valid, it is necessary to prove that the estimation procedure when $ X_{k}=x $ is asymptotically unbiased. For doing so, one needs to prove that the solution of (\ref{eq8}) for any $ \hat{x}\left(0\right)  $ and $ x $ tends to $ x $ as $ t\rightarrow \infty $.

The approximation for the mean error can be written as

\begin{equation}\label{eq49}
\epsilon = \hat{x}-x
\end{equation}

\noindent and the ODE for the mean error is

\begin{equation}\label{eq50}
\frac{d\epsilon}{dt}=\gamma \tilde{h}\left(\epsilon\right),
\end{equation}

\noindent where $ \tilde{h}\left(\epsilon\right)=h\left(\epsilon+x\right)  $ is a function that does not depend on $ x $. 

It is necessary to prove that $ \epsilon \rightarrow 0 $ as $ t\rightarrow \infty $ for every $ \epsilon\left(0\right) \in \mathbb{R} $, which means that $ \epsilon=0 $ is a globally asymptotically stable point \cite{Khalil1992}. Global asymptotic stability of $ \epsilon=0 $ can be shown using an asymptotic stability theorem for nonlinear ODEs. This will require the definition of an unbounded Lyapunov function of the error. To simplify, a quadratic function will be used:

\begin{equation}\label{eq51}
\mathcal{L}\left(\epsilon\right) = \epsilon^{2},
\end{equation}

\noindent which is a positive definite function and tends to infinity when $ \epsilon $ tends to infinity.

If $ \gamma \tilde{h}\left(\epsilon\right) = 0 $ for $ \epsilon=0 $ and $ \frac{d\mathcal{L}}{dt}<0 $ for $ \epsilon\neq 0 $ then by the Barbashin--Krasovskii theorem \cite[Ch. 4]{Khalil1992}, $ \epsilon=0 $ is a globally asymptotically stable point.

To show that both conditions are met, expression (\ref{eq10}) can be rewritten using A6:

\begin{equation}\label{eq52}
h\left(\epsilon\right)=\sum\limits_{i=1}^{\frac{N_{I}}{2}} \eta_{i} \left[ \tilde{F}_{d}\left(i,\epsilon\right)-\tilde{F}_{d}\left(-i,\epsilon\right)\right]  , 
\end{equation}

\noindent where $ \tilde{F}_{d}\left(i,\epsilon\right)=F_{d}\left(i,\epsilon+x,x\right) $ is also a function that does not depend on $ x $.

When $ \epsilon=0 $, the differences between $ \tilde{F}_{d} $ in the sum are differences between probabilities on symmetric intervals, the symmetry of the noise PDF stated in A3 and the symmetry of the quantizer stated in A5 imply that  $ \tilde{h}\left(0\right)=0 $, fulfilling the first condition.

The second condition can be written in more detail by using the chain rule for the derivative:

\begin{equation}\label{eq53}
\frac{d\mathcal{L}}{dt}=\frac{d \mathcal{L}}{ d \epsilon}\frac{d \epsilon}{dt}=2\epsilon \gamma \tilde{h}\left(\epsilon \right) < 0,\quad\text{for}\quad \epsilon\neq 0.
\end{equation}

\noindent As $ \gamma>0 $ by definition, $ \tilde{h}\left(\epsilon\right) $ has to respect the following constraints:
\begin{eqnarray}\label{eq54}
\tilde{h}\left(\epsilon\right)>0,&\text{for $\epsilon < 0$,}\nonumber\\
\tilde{h}\left(\epsilon\right)<0,&\text{for $\epsilon > 0$.}
\end{eqnarray}

When $ \epsilon\neq 0 $, the terms in the sum that gives $ \tilde{h}\left(\epsilon\right) $ are the difference between integrals of the noise PDF under the same interval size but with asymmetric interval centers. Using the symmetry assumptions, for $ \epsilon>0 $, $ \tilde{F}_{d}\left(i,\epsilon\right)  $ is the integration of $ f $ over an interval more distant to zero than for $ \tilde{F}_{d}\left(-i,\epsilon\right)  $, then by the decreasing assumption on $ f $, $ \tilde{F}_{d}\left(i,\epsilon\right)<\tilde{F}_{d}\left(-i,\epsilon\right)  $ and consequently $ \tilde{h}\left(\epsilon\right)<0 $. Using the same reasoning for $ \epsilon<0 $ one can show that $ \tilde{h}\left(\epsilon\right)>0 $. Therefore, the inequalities in (\ref{eq54}) are verified and $ \frac{d\mathcal{L}}{dt}<0  $ for $ \epsilon\neq 0 $.

Finally, as both conditions are satisfied one can say that $ \epsilon=0 $ is globally asymptotically stable, which means that the estimator is asymptotically unbiased and that all the performance results obtained are valid.

Note that from A3 and A5, $ h_{\hat{x}}\left(x\right)<0 $, thus the supplementary condition for stationarity (\ref{eq20}) is also respected.

\subsection{Optimal quantizer parameters}

The performance of the adaptive algorithm can be maximized by minimizing $ \sigma_{\infty}^{2} $ w.r.t. the quantizer levels $ \eta_{i} $. Using (\ref{eq14}) and (\ref{eqhx}) in (\ref{eq18}) gives the following minimization problem:

\begin{equation}\label{eq30}
\arg \underset{\boldsymbol{\eta}}{\min} \left\lbrace \frac{R }{h_{\hat{x}}^{2}}\right\rbrace =\arg \underset{\boldsymbol{\eta}}{\min} \left\lbrace \frac{\boldsymbol{\eta}^{T}\mathbf{F_{d}}\boldsymbol{\eta}}{2\left[ \boldsymbol{\eta}^{T}\mathbf{f_{d}}\right] ^{2}}\right\rbrace ,
\end{equation}

\noindent where $ \boldsymbol{\eta} $ is a vector with the coefficients

\begin{equation}\label{eq31}
\boldsymbol{\eta}=\left[\eta_{1}\, \ldots \,\eta_{\frac{N_{I}}{2} }\right]^{T}.
\end{equation}

\noindent $ \mathbf{F_{d}} $ is a diagonal matrix given by

\begin{equation}\label{eq32}
\mathbf{F_{d}}=\mathrm{diag}\left[ F_{d}\left[1\right],\cdots,F_{d}\left[\frac{N_{I}}{2}\right]\right]
\end{equation}

\noindent and $ \mathbf{f_{d}} $ is the following vector

\begin{equation}\label{eq33}
\mathbf{f_{d}}=\left[f_{d}\left[1\right]\,\cdots\,f_{d}\left[\frac{N_{I}}{2}\right]\right]^{T}.
\end{equation}

The minimization problem is equivalent to the following maximization problem:

\begin{equation}\label{eq34}
\arg \underset{\boldsymbol{\eta}}{\max} \left\lbrace \frac{\left[ \boldsymbol{\eta}^{T}\mathbf{f_{d}}\right] ^{2}}{\boldsymbol{\eta}^{T}\mathbf{F_{d}}\boldsymbol{\eta}}\right\rbrace.
\end{equation}

Using the fact that $ \mathbf{F_{d}} $ is diagonal with non zero diagonal elements, (\ref{eq34}) becomes

\begin{equation}\label{eq35}
\arg \underset{\boldsymbol{\eta}}{\max} \left\lbrace \frac{\left[ \left(\mathbf{F_{d}}^{\frac{1}{2}} \boldsymbol{\eta}\right) ^{T}\left(\mathbf{F_{d}}^{-\frac{1}{2}} \mathbf{f_{d}}\right) \right] ^{2}}{  \left( \mathbf{F_{d}}^{\frac{1}{2}}\boldsymbol{\eta}\right)^{T} \left( \mathbf{F_{d}}^{\frac{1}{2}}\boldsymbol{\eta}\right) }\right\rbrace,
\end{equation}

\noindent the matrices $ \mathbf{F_{d}}^{\frac{1}{2}} $ and $ \mathbf{F_{d}}^{-\frac{1}{2}} $ are obtained by taking the square root and the inverse of the square root of the diagonal elements in $ \mathbf{F_{d}} $. Using the Cauchy--Schwarz inequality on the expression in the numerator gives

\begin{equation}\label{eq36}
\left\lbrace \frac{\left[ \left(\mathbf{F_{d}}^{\frac{1}{2}} \boldsymbol{\eta}\right) ^{T}\left(\mathbf{F_{d}}^{-\frac{1}{2}} \mathbf{f_{d}}\right) \right] ^{2}}{  \left( \mathbf{F_{d}}^{\frac{1}{2}}\boldsymbol{\eta}\right)^{T} \left( \mathbf{F_{d}}^{\frac{1}{2}}\boldsymbol{\eta}\right) }\right\rbrace \leq \mathbf{f_{d}}^{T}\mathbf{F_{d}}^{-1}\mathbf{f_{d}}
\end{equation}

\noindent and the equality happens for

\begin{equation}\label{eq37}
\mathbf{F_{d}}^{\frac{1}{2}}\boldsymbol{\eta} \propto \mathbf{F_{d}}^{-\frac{1}{2}} \mathbf{f_{d}}.
\end{equation}

\noindent Therefore, the optimal $ \boldsymbol{\eta} $ can be chosen to be

\begin{equation}\label{eq38}
\boldsymbol{\eta}^{\star} = \mathbf{F_{d}}^{-1}\mathbf{f_{d}}.
\end{equation}

It is possible to see that the coefficients chosen in this way still depends on $ \Delta $ and $ \boldsymbol{\tau} $. The minimum $ \sigma_{\infty}^{2} $ is

\begin{equation}\label{eq39}
\sigma_{\infty}^{2}=\frac{1}{2\left( \mathbf{f_{d}}^{T}\mathbf{F_{d}}^{-1}\mathbf{f_{d}}\right)} =\left(2 \sum\limits_{i=1}^{\frac{N_{I}}{2}} \frac{f_{d}^{2}\left[i\right]}{F_{d}\left[i\right]}\right) ^{-1}.
\end{equation}

To simplify the choice of the constant $ \Delta $, it will be considered that the noise CDF is parametrized by a known scale parameter $ \delta $, which means that

\begin{equation}\label{deltapar}
F\left(x\right)=F_{n}\left(\frac{x}{\delta}\right),
\end{equation}

\noindent where $ F_{n} $ is the noise CDF for $ \delta=1 $. Thus, the evaluation of the quantizer output levels can be simplified by setting:

\begin{equation}\label{eq42}
\Delta=c_{\Delta}\delta .
\end{equation}

Since the coefficients $ \boldsymbol{\eta}^{\star} $ do not depend on $ x $ anymore, for a given $ c_{\Delta} $ and noise CDF, they can be pre-calculated and stored in a table. For $ i>0 $, these coefficients are given by

\begin{equation}\label{eq43}
\eta_{i}^{\star}=\frac{f_{d}\left[i\right]}{F_{d}\left[i\right]}.
\end{equation}

\noindent Note that for $ \Delta $ given by (\ref{eq42}), $ \eta_{i} $ depends on $ \delta $ only through a $ \frac{1}{\delta} $ multiplicative factor, the other factor can be written as a function of normalized PDFs and CDFs, thus this factor can be pre-calculated based only on the normalized distribution. Note also that the $ \eta_{i}^{\star} $ are given by the score function for estimating a constant location parameter when considering that the offset is fixed and placed exactly at $ x $, therefore this algorithm is equivalent to a gradient ascent technique to maximize the log-likelihood that iterates only one time per observation and sets the offset each time at the last estimate.

Using the $ \eta_{i} $ from (\ref{eq43}), the adaptive estimator can be written as

\begin{equation}\label{eq44}
\hat{X}_{k}=\hat{X}_{k-1}+\gamma_{k}\sign\left(i_{k}\right) \eta_{\left\vert i_{k} \right\vert},
\end{equation}

\noindent with $ i_{k}= Q\left(\frac{Y_{k}-\hat{X}_{k-1}}{\Delta}\right)$.

The sum in (\ref{eq39})  is the Fisher information $ I_{q} $ for estimating a constant $ x $ from the output of the adjustable quantizer with an offset exactly placed at $ x $:

\begin{equation}\label{eq46}
I_{q}=2\sum\limits_{i=1}^{\frac{N_{I}}{2}} \frac{f_{d}^{2}\left[i\right]}{F_{d}\left[i\right]},
\end{equation}

\noindent this quantity can be maximized w.r.t. $ \boldsymbol{\tau} $, thus leading to the following optimization problem:

\begin{equation}\label{eq47}
\boldsymbol{\tau}^{\star}= \underset{\boldsymbol{\tau}}{\arg\max} \quad I_{q}.
\end{equation}

Problem (\ref{eq47}) without constraints on the thresholds seems to be very difficult to solve analytically and no simple solutions for this problem were found in the literature. Therefore, general solutions for (\ref{eq47}) will not be treated here, for the results that will be presented in section V it will be considered that the quantizer is uniform, with $ \boldsymbol{\tau} $ defined as follows

\begin{equation}\label{eq75}
\boldsymbol{\tau}=\left[\tau_{1}= 1 \quad \cdots\quad\tau_{\frac{N_{I}}{2}-1}=\frac{N_{I}}{2}-1\quad\tau_{\frac{N_{I}}{2}}=\infty \right]^{T},
\end{equation}

\noindent then in this case, only $ c_{\Delta} $ need to be set and consequently a grid method can be used.

In the next section the results for each case using the choice of parameters obtained above will be detailed and discussed.

\section{Results and simulation}

It will be supposed that the noise CDF and $ \delta $ are known and also the type of evolution model for $ X_{k} $. Thus for a given $ N_{I} $, $ c_{\delta} $ and $ \boldsymbol{\tau} $, the coefficients $ \eta_{i} $ used in the estimation algorithm (\ref{eq44}) can be calculated using (\ref{eq43}).

There are two quantities that still need to be determined, $ h_{\hat{x}} $ and $ R $. Using (\ref{eq43}) in (\ref{eq14}) and (\ref{eqhx}) gives

\begin{eqnarray}\label{eq55}
h_{\hat{x}}=-2\sum\limits_{i=1}^{\frac{N_{I}}{2}} \frac{f_{d}^{2}\left[i\right]}{F_{d}\left[i\right]}=- I_{q}
\end{eqnarray}
\begin{eqnarray}\label{eq56}
R =2\sum\limits_{i=1}^{\frac{N_{I}}{2}} \frac{f_{d}^{2}\left[i\right]}{F_{d}\left[i\right]}= I_{q}.
\end{eqnarray}

The specific gain $ \gamma_{k} $ and the performance of the algorithm for each model will now be determined.

\subsection{Constant $ X_{k} $}

Replacing $ h_{\hat{x}} $ given by (\ref{eq55}) in (\ref{eq17}) and the result in (\ref{eq7}) gives the following gains:

\begin{eqnarray}\label{eq57}
\gamma_{k}=\frac{1}{k I_{q}}
\end{eqnarray}

\noindent and by replacing (\ref{eq55}) and (\ref{eq56}) in (\ref{eq18}), $ \sigma_{\infty}^{2} $ is obtained:

\begin{eqnarray}\label{eq58}
\sigma_{\infty}^{2} = \frac{1}{I_{q}}.
\end{eqnarray}

In practice this means that for large $ k $, the estimation variance will be (\textit{cf.} (\ref{eq12}))

\begin{eqnarray}\label{eq59}
\var \left[\hat{X}_{k} \right]  \approx \frac{1}{k I_{q}}.
\end{eqnarray}

The right hand side of (\ref{eq59}) is the inverse of the Fisher information for estimating $ X_{k}=x $ based on $ i_{k} $  when the offset is fixed to be $ x $. The inverse of the Fisher information is known as the Cram\'er--Rao bound and it is a lower bound on the variance of unbiased estimators \cite[Ch. 3]{Kay1993}. This means that for large $ k $, the estimator has the lowest possible variance within the class of unbiased estimators using quantized observations with offset $ b_{k}=x $.

In the continuous case (infinite number of quantization intervals) the CRB for $ k $ observations is given by

\begin{eqnarray}\label{eq60}
\mathrm{CRB}_{c}=\frac{1}{k I_{c}},
\end{eqnarray}

\noindent where $ I_{c} $ is the Fisher information given by

\begin{eqnarray}\label{eqIc}
I_{c}=\int \! \left( \frac{f'\left(x\right)}{f\left(x\right) }\right) ^{2} f\left(x\right) \, \mathrm{d}x
\end{eqnarray}

\noindent and $ f'\left(x\right)=\frac{df\left(x\right) }{dx} $. In the cases where $ I_{c} $ exists and for large $ k $, one can calculate the loss of estimation performance $ L_{q} $ in decibels (dB) in the following way:

\begin{eqnarray}\label{eq61}
L_{q}=-10 \log_{10}\left(\frac{\var \left[\hat{X}_{k} \right]}{\mathrm{CRB}_{c}} \right)=-10 \log_{10}\left(\frac{I_{q}}{I_{c}} \right).
\end{eqnarray}

\subsection{Wiener process}

Using (\ref{eq56}) in (\ref{eq22}), the following constant gain is obtained:

\begin{eqnarray}\label{eq60}
\gamma^{\star}=\frac{\sigma_{w}}{\sqrt{I_{q}}}
\end{eqnarray}

\noindent and for this gain, the asymptotic MSE is obtained by substituting (\ref{eq58}) in (\ref{eq25}):

\begin{eqnarray}\label{eq61}
\mathrm{MSE}_{\infty}=\frac{\sigma_{w}}{\sqrt{ I_{q}}}.
\end{eqnarray}

The comparison with the continuous case can be done also using a lower bound on the variance. In this case as $ X_{k} $ is random the Bayesian Cram\'er--Rao bound (BCRB) can be used, this bound is defined as the inverse of the Bayesian information for time $ k $ \cite[Ch. 1]{VanTrees2007}:

\begin{eqnarray}\label{eq62}
\mathrm{BCRB}_{k}=\frac{1}{J_{k}}.
\end{eqnarray}

\noindent For a Wiener process, the Bayesian information can be calculated recursively. The recursive expression, given in its general form in \cite{Tichavsky1998}, for a scalar Wiener process observed with additive noise is

\begin{eqnarray}\label{eq63}
J_{k}=I_{c}+\frac{1}{\sigma_{w}^{2}}-\frac{1}{\sigma_{w}^{4}\left(J_{k-1}+\frac{1}{\sigma_{w}^{2}}\right) }.
\end{eqnarray}

The comparison must be done for $ k\rightarrow \infty $. After calculating the fixed point $ J_{\infty} $ of (\ref{eq63}), the asymptotic BCRB obtained is

\begin{eqnarray}\label{eq64}
\mathrm{BCRB}_{\infty}=\frac{2}{I_{c}+\sqrt{I_{c}^{2}+4\frac{I_{c}}{\sigma_{w}^{2} }}}.
\end{eqnarray}

Expression (\ref{eq61}) is only valid for small $ \sigma_{w} $, in this case (\ref{eq64}) can be approximated by

\begin{eqnarray}\label{eq65}
\mathrm{BCRB}_{\infty}\approx\frac{\sigma_{w}}{\sqrt{I_{c}}}
\end{eqnarray}

\noindent and the loss in asymptotic performance $ L_{q}^{W} $ for the estimation of the Wiener process can be approximated by a function of $ L_{q} $:

\begin{eqnarray}\label{eq66}
L_{q}^{W}\approx\frac{1}{2}L_{q}.
\end{eqnarray}

\subsection{Wiener process with drift}

The varying optimal gain and the MSE are obtained by replacing (\ref{eq55}) and (\ref{eq56}) in (\ref{eq27}) and (\ref{eq28}):

\begin{equation}\label{eq67}
\gamma_{k}^{\star}=\left[\frac{4 u_{k}^{2}} { I_{q}^{2}}\right] ^{\frac{1}{3}}
\end{equation}

\begin{equation}\label{eq68}
\mathrm{MSE}_{k}\approx 3\left[\frac{u_{k}}{4 I_{q}} \right]^{\frac{2}{3}}.
\end{equation}

 As $ u_{k} $ is unknown, it might be estimated. For slowly varying $ u_{k} $ it can be estimated by smoothing the differences between successive estimates:
 
\begin{equation}\label{eq69}
\hat{U}_{k}=\hat{U}_{k-1}+\gamma_{k}^{u}\left[\left( \hat{X}_{k}-\hat{X}_{k-1}\right) - \hat{U}_{k-1}\right] .
\end{equation}

\noindent Then, $ \hat{U}_{k} $ can replace $ u_{k} $ in the evaluation of the gain and the MSE. If more information about the evolution of $ u_{k} $ is known, it might be incorporated in (\ref{eq69}) to have more precise estimates and get closer to the optimal adaptive gain.

As it is hard to have a bound on performance for the estimation of a deterministic signal under non Gaussian noise, the comparison with the continuous observation case will be done using the approximate performance for a nonlinear adaptive algorithm using continuous observations. The algorithm has the following form:

\begin{equation}\label{eq70}
\hat{X}_{k}=\hat{X}_{k-1}+\gamma_{k}^{c}\eta_{c}\left(Y_{k}-\hat{X}_{k-1}\right),
\end{equation}

\noindent where $ \gamma_{k}^{c} $ and the non linearity $ \eta_{c}\left(x\right)  $ are optimized to minimize the MSE.

Using the same theory described for the quantized case it is possible to show that the optimal $ \gamma_{k}^{c} $ and $ \eta_{c}\left(x\right)  $ are

\begin{equation}\label{eq71}
\gamma_{k}^{c}=\left[\frac{4 u_{k}^{2}} { I_{c}^{2}}\right] ^{\frac{1}{3}}
\end{equation}

\begin{equation}\label{eq72}
\eta_{c}\left(x\right)=\frac{f'\left(x\right) }{f\left(x\right) },
\end{equation}

\noindent which exist under the constraint that $ I_{c} $ converges and is not zero and that $ f'\left(x\right) $  exists for every $ x $.

The MSE can be approximated in a similar way as before:

\begin{equation}\label{eq73}
\mathrm{MSE}_{k}\approx 3\left[\frac{u_{k}}{4 I_{c}}\right]^{\frac{2}{3}} .
\end{equation}

Therefore, the loss in performance incurred by quantizing the observations in the estimation of the Wiener process with drift $ L_{q}^{WD} $ can be approximated by

\begin{eqnarray}\label{eq74}
L_{q}^{WD}\approx\frac{2}{3}L_{q}.
\end{eqnarray}

The losses for the three models of $ X_{k} $ depend directly on $ L_{q} $, thus $ L_{q} $ allows to approximate how much of performance is lost for a specific type of noise and threshold set comparing to the optimal (possibly suboptimal in the case with drift) estimator based on continuous measurements. In the next subsection the loss will be evaluated for two different classes of noise considering that the quantization is uniform, then the adaptive algorithm will be simulated in the three cases and the simulated loss will be compared to the results given above to check their validity.

\subsection{Simulation}

The thresholds are considered to be uniform and given by (\ref{eq75}). For a given type of noise, supposing that $ \delta $ is known and for fixed $ N_{I} $, $ I_{q} $ can be evaluated by replacing (\ref{eq75}) and (\ref{eq42}) in the expressions for $ f_{d} $ and $ F_{d} $. As $ I_{q} $ is now a function of $ c_{\Delta} $ only, it can be maximized by adjusting this parameter. Being a scalar maximization problem this can be done by using grid optimization (searching for the maximum in a fine grid of possible $ c_{\Delta} $). After finding the optimal $ c_{\Delta} $ and $ I_{q} $, the coefficients $ \eta_{i} $, the optimal gains $ \gamma_{k} $ and the quantizer input gain $ \frac{1}{\Delta} $ can be evaluated and then all the parameters are defined.

Note that it is supposed that the model for $ X_{k} $ is known as setting $ \gamma_{k} $ depends on it. As a consequence of this assumption, in a real application the choice between the three models must be clear. When this choice is not clear from the application, it is always simpler to choose $ X_{k} $ to be a Wiener process, first, because the complexity of the algorithm is lower and second, because supposing that the increments are Gaussian and i.i.d. does not impose too much information on the evolution of $ X_{k} $. Still, $ \sigma_{w} $ must be known, in practice it can be set based on prior knowledge on the possible variation of $ X_{k} $ or by accepting a slower convergence and a small loss of asymptotic performance, it can be estimated jointly with $ X_{k} $ using an extra adaptive estimator for it. In the last case, when it is known that the increments of $ X_{k} $ have a deterministic component, the fact the $ \gamma_{k} $ depends on $ u_{k} $ is not very useful and prior information on the variations of $ X_{k} $ are not normally as detailed as knowing $ u_{k} $ itself, making it necessary to accept a small loss of performance to estimate $ u_{k} $ jointly. The estimation of $ u_{k} $ can be done using (\ref{eq69}) where prior knowledge on the variations of $ u_{k} $ can be integrated in the gain $ \gamma_{k}^{u} $. If precise knowledge on the evolution of $ u_{k} $ is known through dynamical models, then it might be more useful to use other forms of adaptive estimators known as multi-step algorithms \cite[Ch. 4]{Benveniste1990}.

The evaluation of the loss and the verification of the results will be done considering two different classes of noise that verify assumptions A1 to A3, namely, generalized Gaussian (GG) noise and Student's-t (ST) noise. The motivation for the use of these two densities comes from signal processing, statistics and information theory.

In signal processing, when additive noise is not constrained to be Gaussian a common assumption is that the noise follows a GG distribution \cite{Varanasi1989}. This distribution not only contains the Gaussian case as an specific example, but also by changing one of its parameters, one can represent from the impulsive Laplacian case to distributions close to the uniform case. In robust statistics, when the additive noise is considered to be impulsive, a general class for the distribution of the noise is the ST distribution \cite{Lange1989}. ST distribution includes as a specific case the Cauchy distribution, known to be heavy tailed and thus normally used in robust statistics, also by changing a parameter of the distribution an entire class of heavy tailed distributions can be represented. When looking from an information point of view, if no priors on the noise distributions are given, noise models must be as random as possible to ensure that the noise is an uninformative part of the observation, thus noise models must maximize some criterium of randomness. Commonly used criteria for randomness are entropy measures and both distributions considered above are entropy maximizers. GG distributions maximize the Shannon entropy under constraints on the moments \cite[Ch. 12]{Cover2006} and ST distributions maximize the R\'enyi entropy under constraints on the second order moment \cite{Costa2003}.

Both distributions are parametrized by a shape parameter $ \beta \in \mathbb{R}^{+} $ and their PDFs and CDFs for $ \delta=1 $  are

\begin{eqnarray}\label{eq76}
f_{GG}\left(x\right) &=& \frac{\beta}{2\Gamma\left(\frac{1}{\beta}\right) }\expon^{-\left\vert x \right\vert^{\beta}}, \\
F_{GG}\left(x\right) &=& \frac{1}{2}\left[1+\sign\left(x\right)\frac{\gamma\left(\frac{1}{\beta},\left\vert x \right\vert^{\beta} \right) }{\Gamma\left(\frac{1}{\beta} \right) }  \right],
\end{eqnarray}

\noindent for the GG distribution, where $ \gamma\left( \cdot,\cdot \right)  $ is the incomplete gamma function and $ \Gamma\left( \cdot \right)  $ is the gamma function,

\begin{eqnarray}\label{eq77}
f_{ST}\left(x\right) &=& \frac{\Gamma\left( \frac{\beta+1}{2}\right) }{\sqrt{\beta \pi} \Gamma\left(\frac{\beta}{2} \right) } \left(1+\frac{1}{\beta}x^{2}\right)^{-\frac{\beta +1}{2}} , \\
F_{ST}\left(x\right) &=&  \frac{1}{2}\left\lbrace 1+\sign\left(x\right)\left[1-I_{\frac{\beta}{x^{2}+\beta}}\left(\frac{\beta}{2},\frac{1}{2}\right) \right]   \right\rbrace,\quad
\end{eqnarray}

\noindent for the ST distribution, where $ I_{\frac{\beta}{x^{2}+\beta}}\left( \cdot,\cdot\right)  $ is the incomplete beta function.

\subsubsection{\textbf{Performance loss} - $ L_{q} $}

The first quantity to be evaluated will be the loss $ L_{q} $. To evaluate $ L_{q} $, after evaluating $ I_{q} $ based on $ f $ and $ F $ defined above, it is also needed to evaluate $ I_{c} $. Evaluating the integral on (\ref{eqIc}), one obtains for the GG and ST distributions respectively:

\begin{eqnarray}\label{eq78}
I_{GG}\left(x\right) &=&  \frac{\beta \left(\beta-1\right)\Gamma\left(1-\frac{1}{\beta} \right)  }{\Gamma\left(\frac{1}{\beta}\right) }, \\
I_{ST}\left(x\right) &=&  \frac{\beta+1}{\beta+3}.
\end{eqnarray}

\begin{figure}[!t]
\centering
\includegraphics{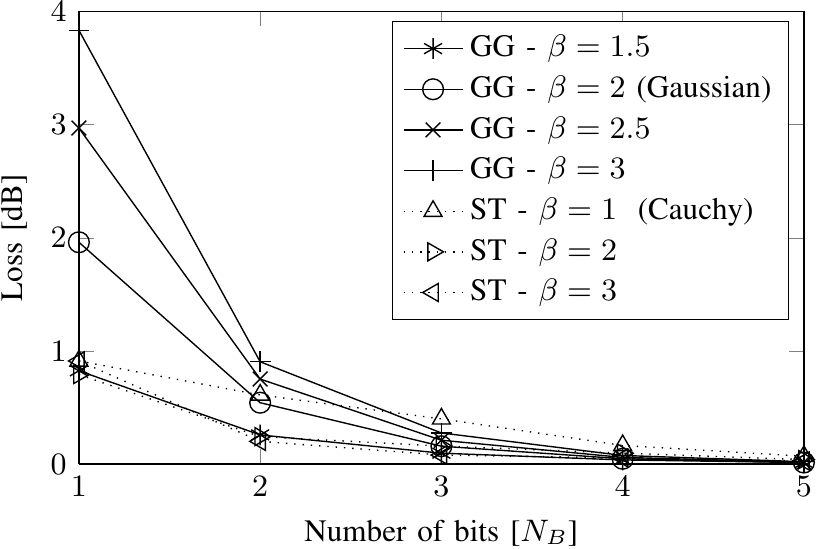}
\caption{Loss of performance due to quantization of measurements for different types of noise and number of quantization bits.}
\label{fig3}
\end{figure}

The loss was evaluated for $ N_{I}=\left\lbrace 2,4,8,16,32 \right\rbrace  $ which corresponds to $ N_{B}=\log_{2}\left(N_{I}\right) =\left\lbrace  1,2,3,4,5\right\rbrace $ number of bits and for the shape parameters $ \beta=\left\lbrace 1.5,2,2.5,3 \right\rbrace  $ for GG noise and $ \beta=\left\lbrace 1,2,3 \right\rbrace  $ for ST noise. The results are shown in Fig. \ref{fig3}. As it was expected, the loss reduces with increasing $ N_{B} $. It is interesting to note that the maximum loss, observed for $ N_{B}=1 $, goes from approximately $ 1\text{dB} $ to $ 4\text{dB} $, which represents factors less than 3 in MSE increase for estimating a constant with 1 bit quantization. Also interesting is the fact that the loss decreases rapidly with $ N_{B} $, for 2 bits quantization all the tested types of noise produce losses below $ 1\text{dB} $, resulting in linear increases in MSE not larger than 1.3. This indicates that when using the adaptive estimators developed here, it is not very useful to use more than 4  or 5 bits for quantization.

\begin{figure*}[!ht]
\subfloat[]{\includegraphics{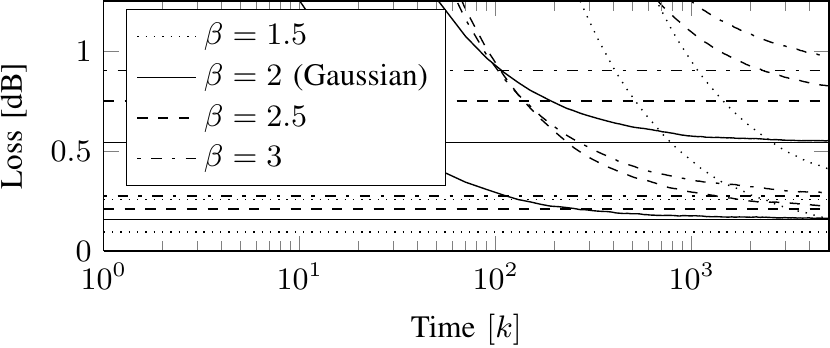} \label{fig4a}} 
\subfloat[]{\includegraphics{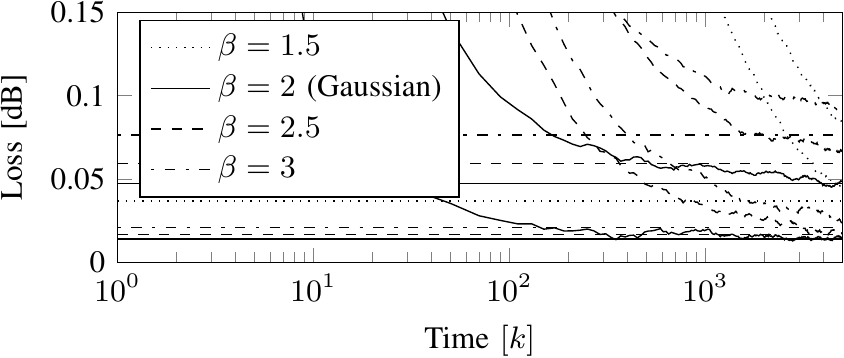} \label{fig4b}}\\
\subfloat[]{\includegraphics{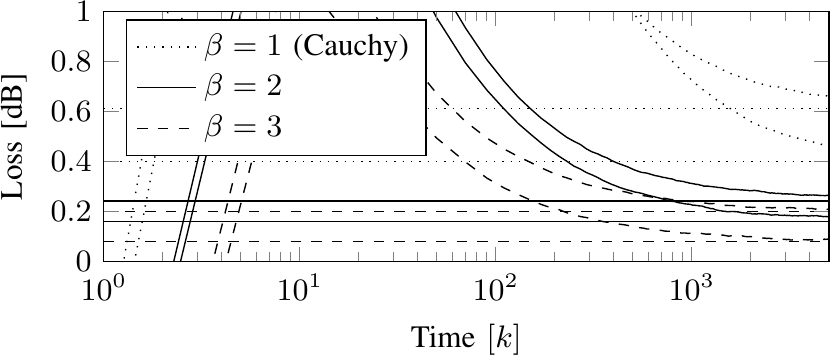} \label{fig4c}}
\subfloat[]{\includegraphics{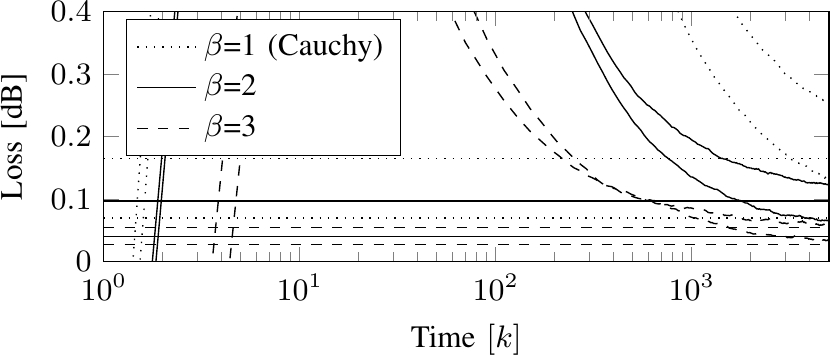} \label{fig4d}}
\caption{\textbf{Constant.} Quantization loss of performance for GG and ST noises and $ N_{B}=\left\lbrace 2,3,4,5\right\rbrace  $ when $ X_{k} $ is constant. For each type of noise there are 4 curves, the constant losses are the theoretical results and the decreasing losses are the simulated results, thus producing pairs of curves of the same type, for each pair the higher results represent lower number of quantization bits. In (a) results for GG noise and $ N_{B}=2 $ and $ 3 $, in (b) the results for GG noise and $ N_{B}=4 $ and $ 5 $ are shown. The figures (c) and (d) are the results for ST noise, in (c) $ N_{B}=2 $ and $ 3 $ are considered while in (d) $ N_{B}=4 $ and $ 5 $.}
\label{fig4}
\end{figure*}

The performance for 2 bits seems to be related to the noise tail, note that smaller losses were obtained for distributions with heavier tail (ST distributions and GG distribution with $ \beta=1.5 $),
this is due to the fact that for large tail distributions a small region around the median of the distribution is very informative, thus as most of the information is contained there, when the only threshold available is placed there, the relative gain of information is greater than in the other cases, leading to smaller losses. This can also be the reason for the slow decrease of the loss for these distributions, as the quantizer thresholds are placed uniformly, some of them will be placed in the non informative amplitude region and consequently the decrease in loss will be not as sharp as in the other cases.

Laplacian distribution was not tested, because for this distribution the optimal adaptive estimator in the continuous case is already an adaptive estimator with a binary quantizer. This can be seen easily if one evaluates $ I_{q} $ as a function of the thresholds, the result will be a constant for all possible sets of thresholds meaning that they are unimportant, moreover, if $ \eta_{i} $ are evaluated one will find that they are all equal, therefore only the sign of the difference between the observations and the last estimate is important. Consequently, the loss found in this case would be a constant for all $ N_{B} $.

To validate the results, the adaptive algorithms will be simulated and the loss obtained will be compared to the approximations given above. The simulation results will be presented in the same order as before, first the constant case, then the Wiener process case and finally the case with drift. All the simulation were done considering $ N_{B}=\left\lbrace 2,3,4,5\right\rbrace $.

\subsubsection{Simulated loss - \textbf{Constant}}

in the constant case, the 7 types of noise with evaluated $ L_{q} $ were tested, the value of $ X_{0}=x $ was set to be zero and the initial condition of the adaptive algorithm was set with a small error ($ \hat{X}_{1} \in\left\lbrace0,10 \right\rbrace  $), the number of samples was set to be $ 5000 $ to have sufficient points for convergence, the algorithm was simulated $ 2.5\times 10^{6} $ times and the error results were averaged to produce a simulated MSE. Based on the simulated MSE a simulated loss was calculated, and it is shown in Fig. \ref{fig4}.

The simulated results seems to converge to the theoretical approximations of $ L_{q} $, thus validating these approximations. This also means that the variance of estimation tends in simulation to the CRB for quantized observations, validating the fact that the algorithm is asymptotically optimal. The convergence time looks to be related to $ N_{B} $, when $ N_{B} $ increases the time to get closer to the optimal performance decreases.

\begin{figure}[!t]
\centering
\includegraphics{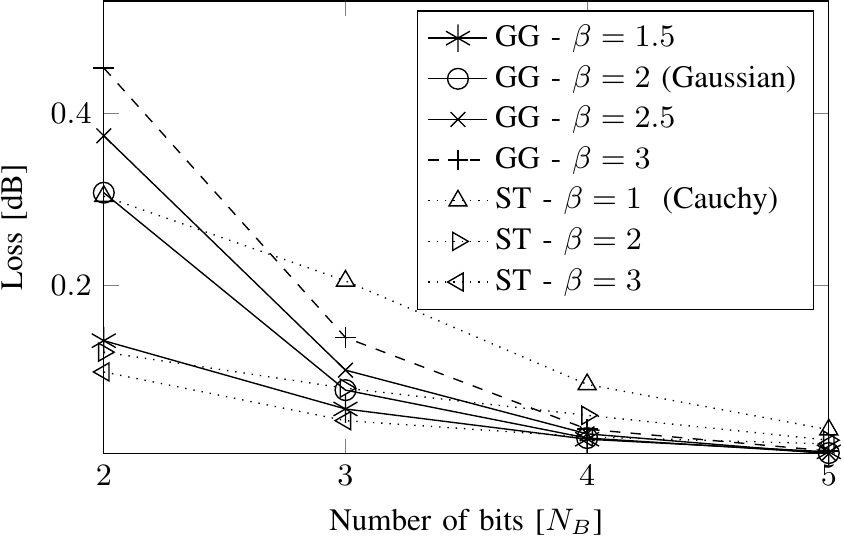}
\caption{\textbf{Wiener process.} Simulated quantization performance loss for a Wiener process $ X_{k} $ with $ \sigma_{w}=0.001 $, different types of noise and number of quantization bits.}
\label{fig5}
\end{figure}

\subsubsection{Simulated loss -\textbf{Wiener process}}

for a Wiener process, $ L_{q} $ was evaluated by setting $ \hat{X}\left(0\right)  $ randomly around 0 and $X_{0}=0$, then $ 10^{4} $ realizations with $ 10^{5} $ samples were simulated and the MSE was estimated by averaging the realizations of the squared error for each instant, then as it was observed that the error was approximately stationary after $ k=1000 $, the sample mean squared error was also averaged resulting in an estimate of the asymptotic MSE. Based on the obtained values of the MSE a simulated loss was evaluated. The results for the 7 types of noise and $ \sigma_{w}=0.001 $ are shown in Fig. \ref{fig5}.

\begin{figure}[!t]
\centering
\includegraphics{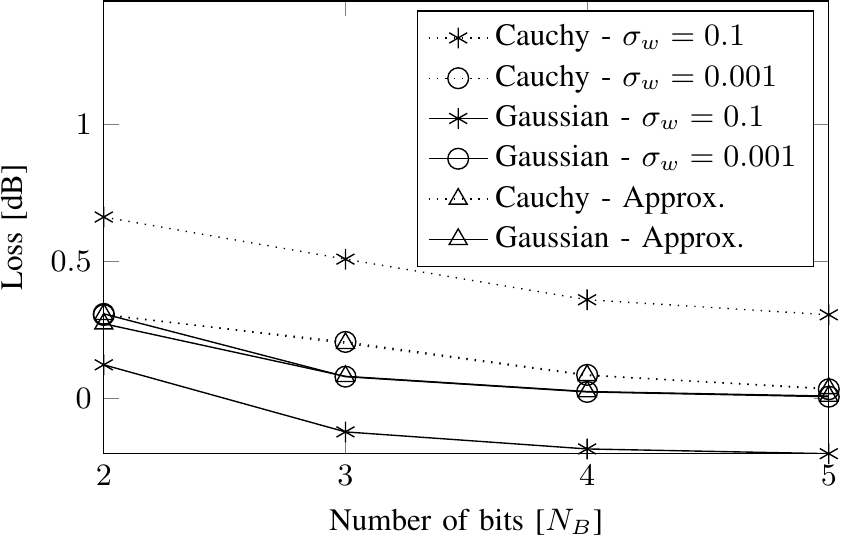}
\caption{\textbf{Wiener process.} Comparison of simulated and theoretical losses in the Gaussian and Cauchy noise cases when estimating a wiener process with $ \sigma_{w}=0.1 $ or $ \sigma_{w}=0.001 $.}
\label{fig6}
\end{figure}

As expected, the results have the same form of the theoretical loss given in Fig. \ref{fig3}. To verify the results for different $ \sigma_w $, the loss was evaluated through simulation also for $ \sigma_w=0.1 $ in the Gaussian (GG with $ \beta=2 $) and Cauchy cases (ST with $ \beta=1 $). The results are shown in Fig. \ref{fig6}, where the theoretical losses for these cases are also shown. It is clear from the results that $ X_{k} $ might move slowly to give a performance close to the theoretical results, but it is also interesting that the simulated loss seems to have the same decreasing rate as a function of $ N_{B} $ when compared to the theoretical results. This means that the dependence on $ I_{q} $ of the MSE seems to still be correct and it indicates that even in a faster regime for $ X_{k} $, the thresholds can be set by maximizing $ I_{q} $.

\begin{figure}[!t]
\centering
\includegraphics{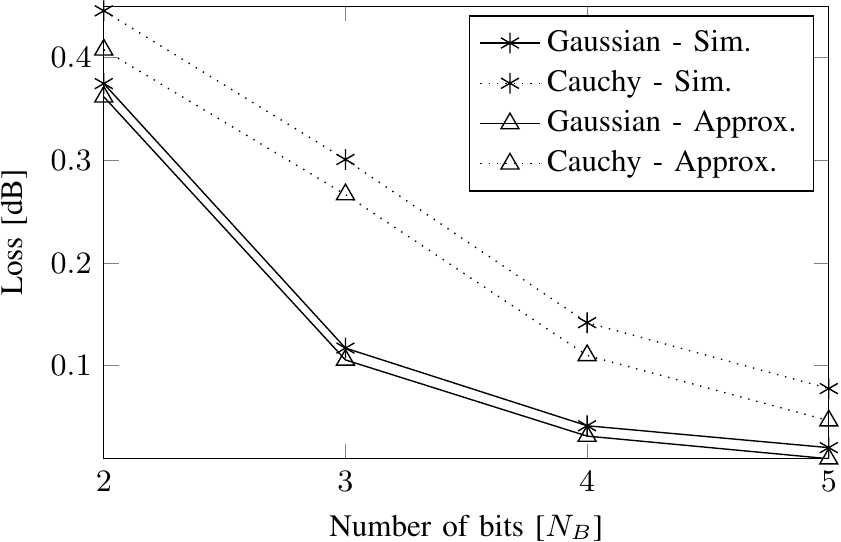}
\caption{\textbf{Wiener process with drift.} Comparison of simulated and theoretical losses in the Gaussian and Cauchy noise cases for estimating a Wiener process with constant mean drift $ u_{k}=10^{-4} $ and standard deviation $ \sigma_{w}=10^{-4} $.}
\label{fig7}
\end{figure}

\subsubsection{Simulated loss - \textbf{Wiener process with drift}}

for $ X_{k} $ with drift, $ W_{k} $ was simulated with mean and standard deviations $ u_{k}=\sigma_{w}=10^{-4} $, which represents a slow linear drift with small random fluctuations, the initial conditions were set to be $ X_{0}=\hat{X}=0 $ and the drift estimator was set with constant gain $\gamma_{k}^{u} =10^{-5} $. Its initial condition was set to the true $ u_{k} $ to reduce the transient time and consequently the simulation time. As $ u_{k} $ is constant, the loss evaluation was done in the same form as for $ X_{k} $ without drift, based on averaging through realizations and time. The results for the Gaussian and Cauchy cases are shown in Fig. \ref{fig7}.

The small offset between simulated and theoretical results is produced by the joint estimation of $ u_{k} $. Note that keeping $ \gamma_{k}^{u} $ to a small constant allows to adaptively follow slow variations in $ u_{k} $. The convergence to the simulated loss in Fig. \ref{fig7} was also obtained for simulations with errors in the initial conditions but in this case the transient regime was very long, indicating that other schemes might be considered when the theoretical performance is needed in a short period of time. Multi-step adaptive algorithms could be used for faster convergence to the theoretical performance but they would need a precise model for the evolution of the drift which is not considered here.

\section{Conclusions}

In this work an adaptive estimation algorithm based on quantized observations was proposed.  Based on observations with additive noise and quantized with adjustable offset and gain, the objective was to estimate with a low complexity online adaptive algorithm a scalar parameter that could follow one of three models, constant, Wiener process and Wiener process with drift. Under the hypothesis that the noise PDF is symmetric and strictly decreasing, and that quantizer is also symmetric, by using Lyapunov theory it was shown that for the optimal quantizer output coefficients, the algorithm is asymptotically stable. It was also shown that the asymptotic performance in terms of mean squared error could be optimized by using static update coefficients that depend only on the shape of the observation noise and on the quantizer thresholds.

Performance results were obtained based on the optimal choice of the quantizer output levels. It was observed that the effect of quantization on performance could be quantified by the Fisher information of the quantized observations. Thus, this clearly indicates that the quantizer thresholds must be placed to maximize the Fisher information. It was also observed that for the three models, the loss of performance of the algorithm w.r.t. the optimal continuous measurement is given by a function of the ratio of the corresponding Fisher informations.

For testing the results, two different families of noise were considered, generalized Gaussian noise and Student's-t noise, both under uniform quantization. First, the theoretical loss was evaluated for different numbers of quantization intervals. The results indicate that with only a few quantization bits (4 and 5) the adaptive algorithm performance is very close to the continuous observation case and it was observed that uniform quantization seems to penalize more estimation performance under heavy tailed distributions. 

Estimation in the three possible scenarios was simulated and the results validated the accuracy of the theoretical approximations. In the constant case it was observed that the algorithm performance was very close to the Cram\'er--Rao bound, in the Wiener process case it was observed that the theoretical results are very accurate for small increments of the Wiener process and in the drift case it was seen that by accepting a small increase in the mean squared error it is possible to estimate jointly the drift.

Another interesting result is that a varying parameter has a loss of performance smaller than a constant parameter, thus a type of dithering effect seems to be present. In this case, the variations of the input signal makes the tracking performance of the estimator to get close to the continuous measurement performance.

The fact that the number of quantization bits does not influence much the performance of estimation leads to conclude that it seems more reasonable to focus on using more sensors than using high resolution quantizers for increasing performance. Consequently, this motivates the use of sensor network approaches. 

As the Fisher information for quantized measurements plays a central role in the performance of the algorithms, the study of its properties as a function of the noise type and quantizer thresholds seems to be a subject for future work. A possible approach for the study of its general behavior would be to consider high resolution approximations.

Finally, as in practice sensor noise scale parameter and Wiener process increment standard deviation can be unknown and slowly variable, it would be also interesting to study how the algorithm design and performance would change by estimating all these parameters jointly.

% needed in second column of first page if using \IEEEpubid
%\IEEEpubidadjcol

%%%%%%%%%%%%%%%%%%%%%%%%%%%%%%%%%

% use section* for acknowledgement
\section*{Acknowledgment}
The authors would like to thank Eric Moisan, Steeve Zozor and Olivier J. J. Michel for their helpful comments and the Erasmus Mundus EBWII program for funding this study.

%%%%%%%%%%%%%%%%%%%%%%%%%%%%%%%%%%%%%%%%%%%%%%%%%%%%%%%%%%%%%%%%%%

% Can use something like this to put references on a page
% by themselves when using endfloat and the captionsoff option.
\ifCLASSOPTIONcaptionsoff
  \newpage
\fi

\bibliographystyle{IEEEtran}
\IEEEtriggeratref{5}
\bibliography{biblio}

% Generated by IEEEtran.bst, version: 1.13 (2008/09/30)
\begin{thebibliography}{10}
\providecommand{\url}[1]{#1}
\csname url@samestyle\endcsname
\providecommand{\newblock}{\relax}
\providecommand{\bibinfo}[2]{#2}
\providecommand{\BIBentrySTDinterwordspacing}{\spaceskip=0pt\relax}
\providecommand{\BIBentryALTinterwordstretchfactor}{4}
\providecommand{\BIBentryALTinterwordspacing}{\spaceskip=\fontdimen2\font plus
\BIBentryALTinterwordstretchfactor\fontdimen3\font minus
  \fontdimen4\font\relax}
\providecommand{\BIBforeignlanguage}[2]{{%
\expandafter\ifx\csname l@#1\endcsname\relax
\typeout{** WARNING: IEEEtran.bst: No hyphenation pattern has been}%
\typeout{** loaded for the language `#1'. Using the pattern for}%
\typeout{** the default language instead.}%
\else
\language=\csname l@#1\endcsname
\fi
#2}}
\providecommand{\BIBdecl}{\relax}
\BIBdecl

\bibitem{Chong2003}
C.~Chong and S.~Kumar, ``Sensor networks: Evolution, opportunities, and
  challenges,'' \emph{Proceedings of the IEEE}, vol.~91, no.~8, pp. 1247--1256,
  2003.

\bibitem{Gersho1992}
A.~Gersho and R.~Gray, \emph{Vector quantization and signal compression}.\hskip
  1em plus 0.5em minus 0.4em\relax Springer, 1992.

\bibitem{Papadopoulos2001}
H.~Papadopoulos, G.~Wornell, and A.~Oppenheim, ``Sequential signal encoding
  from noisy measurements using quantizers with dynamic bias control,''
  \emph{IEEE Trans. Inf. Theory}, vol.~47, no.~3, pp. 978--1002, 2001.

\bibitem{Ribeiro2006a}
A.~Ribeiro and G.~Giannakis, ``Bandwidth-constrained distributed estimation for
  wireless sensor networks-part {I}: {Gaussian} case,'' \emph{IEEE Trans.
  Signal Process.}, vol.~54, no.~3, pp. 1131--1143, 2006.

\bibitem{Li2007}
H.~Li and J.~Fang, ``Distributed adaptive quantization and estimation for
  wireless sensor networks,'' \emph{IEEE Signal Process. Lett.}, vol.~14,
  no.~10, pp. 669--672, 2007.

\bibitem{Fang2008}
J.~Fang and H.~Li, ``Distributed adaptive quantization for wireless sensor
  networks: From delta modulation to maximum likelihood,'' \emph{IEEE Trans.
  Signal Process.}, vol.~56, no.~10, pp. 5246--5257, 2008.

\bibitem{Benveniste1990}
A.~Benveniste, M.~M{\'e}tivier, and P.~Priouret, \emph{Adaptive algorithms and
  stochastic approximations}.\hskip 1em plus 0.5em minus 0.4em\relax
  Springer-Verlag New York, Inc., 1990.

\bibitem{Khalil1992}
H.~Khalil and J.~Grizzle, \emph{Nonlinear systems}.\hskip 1em plus 0.5em minus
  0.4em\relax Macmillan Publishing Company New York, 1992.

\bibitem{Kay1993}
S.~Kay, \emph{Fundamentals of statistical signal processing, Volume 1:
  Estimation theory}.\hskip 1em plus 0.5em minus 0.4em\relax PTR Prentice Hall,
  1993.

\bibitem{VanTrees2007}
H.~L. Van~Trees and K.~L. Bell, \emph{{Bayesian Bounds for Parameter Estimation
  and Nonlinear Filtering/Tracking}}.\hskip 1em plus 0.5em minus 0.4em\relax
  Wiley-IEEE Press, 2007.

\bibitem{Tichavsky1998}
P.~Tichavsky, C.~Muravchik, and A.~Nehorai, ``Posterior {Cram\'er}--{Rao}
  bounds for discrete-time nonlinear filtering,'' \emph{IEEE Trans. Signal
  Process.}, vol.~46, no.~5, pp. 1386 --1396, 1998.

\bibitem{Varanasi1989}
M.~Varanasi and B.~Aazhang, ``Parametric generalized {Gaussian} density
  estimation,'' \emph{The Journal of the Acoustical Society of America},
  vol.~86, pp. 1404--1415, 1989.

\bibitem{Lange1989}
K.~Lange, R.~Little, and J.~Taylor, ``Robust statistical modeling using the t
  distribution,'' \emph{Journal of the American Statistical Association}, pp.
  881--896, 1989.

\bibitem{Cover2006}
T.~M. Cover and J.~A. Thomas, \emph{Elements of Information Theory 2nd
  Edition}.\hskip 1em plus 0.5em minus 0.4em\relax Wiley-Interscience, 2006.

\bibitem{Costa2003}
J.~Costa, A.~Hero, and C.~Vignat, ``On solutions to multivariate maximum
  $\alpha$-entropy problems,'' in \emph{Energy Minimization Methods in Computer
  Vision and Pattern Recognition}, ser. Lecture Notes in Computer Science,
  A.~Rangarajan, M.~Figueiredo, and J.~Zerubia, Eds.\hskip 1em plus 0.5em minus
  0.4em\relax Springer Berlin/Heidelberg, 2003, vol. 2683, pp. 211--226.

\end{thebibliography}

\begin{IEEEbiographynophoto}{Rodrigo Cabral Farias}
was born in Porto Alegre, Brazil, in 1986. He received the B.Sc. degree in electrical engineering from the Federal University of Rio Grande do Sul (UFRGS), Porto Alegre, Brazil, and from the Grenoble Institute of Technology (Grenoble-INP), Grenoble, France, both in 2009. He received the M.Sc degree in signal processing from the Grenoble-INP in 2009. He is currently pursuing the Ph.D. degree in signal processing at the GIPSA-Lab (Grenoble Laboratory of Image, Speech, Signal, and Automation).

His research concerns statistical signal processing, digital communications and sensor networks.
\end{IEEEbiographynophoto}
\begin{IEEEbiographynophoto}{Jean-Marc Brossier}
was born in Thonon, France, in 1965. He received the Ph.D. degree in signal processing in 1992 and the Habilitation a Diriger des Recherches in 2002, both from Grenoble- INP.

He worked as an Assistant Professor for Saint-Etienne University (Universit\'e Jean Monnet) from 1993 to 1995. Since 1995, he has been with Grenoble-INP and GIPSA-Lab. He is now a Professor of electrical engineering and he lectures on signal processing and digital communications. His research interests include statistical signal processing, digital communications, adaptive algorithms and physics.
\end{IEEEbiographynophoto}

\vfill
% You can push biographies down or up by placing
% a \vfill before or after them. The appropriate
% use of \vfill depends on what kind of text is
% on the last page and whether or not the columns
% are being equalized.

% Can be used to pull up biographies so that the bottom of the last one
% is flush with the other column.
%\enlargethispage{-5in}

\end{document}